# Quantum teleportation via a two-qubit Heisenberg *XXZ* chain
# ——effects of anisotropy and magnetic field

Yue Zhou[1], Guo-Feng Zhang[2*]

[1]*State Key Laboratory for Superlattices and Microstructures, Institute of Semiconductors, Chinese Academy of Sciences, P. O. Box 912, Beijing 100083, People's Republic of China*

[2]*Department of Physics, School of Sciences, Beijing University of Aeronautics & Astronautics, Xueyuan Road No. 37, Beijing 100083, People's Republic of China*

**Abstract:** We study quantum teleportation via a two-qubit Heisenberg *XXZ* chain under an inhomogeneous magnetic field. We first consider entanglement teleportation, and then focus on the teleportation fidelity under different conditions. The effects of anisotropy and the magnetic field, both uniform and inhomogeneous, are discussed. We also find that, though entanglement teleportation does require an entangled quantum channel, a nonzero critical value of minimum entanglement is not always necessary.

## 1. INTRODUCTION

Quantum teleportation is a fascinating phenomenon based on the nonlocal property of quantum mechanics. During teleportation, the unknown teleported state is destroyed in the sender's location and then recurs at the distant receiver's. This process relies only on local unitary transformations and some assistance from classical communication. Theoretically speaking, arbitrary quantum states can be teleported with fidelity (see section IV) better than any classical protocol [1]. As an important component of quantum information, quantum teleportation has received extent investigation both theoretically and experimentally.

Of the many schemes proposed, the scheme based on Heisenberg interaction in solid state systems is an attractive one. In solid state systems, the state at thermal equilibrium (temperature $T$) is described by $\rho(T) = e^{-H/kT}/Z$, where $H$ is the Hamiltonian, $Z = tr e^{-H/kT}$ is the partition function and $k$ is Boltzmann's constant. The entanglement associated with thermal equilibrium state $\rho(T)$ is called thermal entanglement. As potential resources, quantum teleportation via thermally entangled states of Heisenberg spin chains has been investigated by several authors [2-4]. However, the former works devote attention primarily to the Heisenberg *XY* and *XXX* chain, and the anisotropic *XXZ* model has never been considered. The *XXZ* chain is a familiar Heisenberg chain and has been used to describe quantum computers based on NMR [5] and on electrons in Helium [6, 7], so it is worth investigating. Also, the solid state structure is intrinsically inhomogeneous and magnetic imperfections or impurities are likely to be present, which lead to inhomogeneous stray magnetic fields. Thus, it is necessary to take inhomogeneous magnetic fields into account. In this paper we study the quantum teleportation via thermal entanglement of Heisenberg *XXZ* chain in the presence of an inhomogeneous magnetic field. Both the ferromagnetic cases and antiferromagnetic cases are considered.

---

[*] Corresponding author; electronic address: gf1978zhang@buaa.edu.cn

## 2. THE THERMAL ENTANGLEMENT OF TWO-QUBIT HEISENBERG *XXZ* MODEL

The Hamiltonian of a two-qubit Heisenberg *XXZ* chain under an inhomogeneous magnetic field is given by

$$H = \frac{1}{2}\left[J(\sigma_1^x \sigma_2^x + \sigma_1^y \sigma_2^y + \lambda \sigma_1^z \sigma_2^z) + (B+b)\sigma_1^z + (B-b)\sigma_2^z\right], \quad (1)$$

where $B$ is the average magnetic field along the $z$ direction, and parameter $b$ represents the degree of inhomogeneity; $J$ is the exchange coupling in the *x-y* plane, the model is called antiferromagnetic for $J > 0$ and ferromagnetic for $J < 0$; $\lambda > 0$ is the anisotropy in the $z$ direction. The eigenstates and corresponding eigenvalues of the Hamiltonian (1) can be expressed as

$$\varphi_1 = |00\rangle, \qquad E_1 = \frac{1}{2}(\lambda J - 2B),$$

$$\varphi_2 = |11\rangle, \qquad E_2 = \frac{1}{2}(\lambda J + 2B),$$

$$\varphi_3 = \frac{\varepsilon}{\sqrt{J^2 + \varepsilon^2}}|10\rangle + \frac{J}{\sqrt{J^2 + \varepsilon^2}}|01\rangle, \qquad E_3 = -\frac{\lambda J}{2} - \eta,$$

$$\varphi_4 = \frac{\zeta}{\sqrt{J^2 + \zeta^2}}|10\rangle + \frac{J}{\sqrt{J^2 + \zeta^2}}|01\rangle, \qquad E_4 = -\frac{\lambda J}{2} + \eta, \quad (2)$$

where $\eta = \sqrt{b^2 + J^2}$, $\varepsilon = b - \eta$, $\zeta = b + \eta$; $|1\rangle$ and $|0\rangle$ denote the spin-up and spin-down states, respectively.

At the thermal equilibrium (temperature $T$), the density matrix $\rho(T)$ of the above system is given by

$$\rho(T) = \frac{1}{Z}\begin{pmatrix} e^{-E_2/T} & 0 & 0 & 0 \\ 0 & e^{\lambda J/2T}(m-n) & -c & 0 \\ 0 & -c & e^{\lambda J/2T}(m+n) & 0 \\ 0 & 0 & 0 & e^{-E_1/T} \end{pmatrix} \quad (3)$$

in the standard basis { $|11\rangle$, $|10\rangle$, $|01\rangle$, $|00\rangle$ }, where $Z = 2e^{-\lambda J/2T}\cosh(B/T) + 2e^{\lambda J/2T}\cosh(\eta/T)$, $m = \cosh(\eta/T)$, $n = b\sinh(\eta/T)/\eta$ and $c = e^{\lambda J/2T}J\sinh(\eta/T)/\eta$. Note that we set $k = 1$ from hereon for simplification. The thermal entanglement of this model is presented in detail in Ref. [8].

## 3. ENTANGLEMENT TELEPORTATION

As the name implies, entanglement teleportation means to teleport an entangled quantum state. If the quantum channel is maximally entangled, the teleportation will be perfectly achieved. However, for real channels, where the maximal entanglement is generally inaccessible, the entanglement of the teleported states will be more or less lost. In this section, we focus on how much the entanglement is transferred under various conditions.

Without loss of generality, we suppose the input state is $|\varphi\rangle_{in} = \cos\theta|11\rangle + e^{i\phi}\sin\theta|00\rangle$, where $0 \leq \phi < 2\pi$ is the phase difference between the two bases and $0 \leq \theta \leq \pi/2$ describes the amplitude. As is shown below, $\theta$ also controls the degree of entanglement. When the state described by Eq. (3) acts as the quantum channel, the output state is given by [9]

$$\rho_{out} = \sum_{ij} p_{ij}(\sigma_i \otimes \sigma_j)\rho_{in}(\sigma_i \otimes \sigma_j), \tag{4}$$

in which $\sigma_i(i=0,1,2,3)$ denote the unit matrix $I$ and three components of Pauli matrix, respectively; $p_{ij} = tr[E_i\rho(T)] \cdot tr[E_j\rho(T)]$; $E^i = |\psi_i\rangle\langle\psi_i|$ where $|\psi_0\rangle = 2^{-1/2}(|01\rangle - |10\rangle)$, $|\psi_1\rangle = 2^{-1/2}(|00\rangle - |11\rangle)$, $|\psi_2\rangle = 2^{-1/2}(|00\rangle - |11\rangle)$ and $|\psi_3\rangle = 2^{-1/2}(|01\rangle + |10\rangle)$. So that we obtain:

$$\rho_{out} = \frac{1}{Z^2}\begin{pmatrix} a_2\sin^2\theta + a_3\cos^2\theta & 0 & 0 & \frac{1}{2}a_4\sin(2\theta)e^{-i\phi} \\ 0 & a_1 & 0 & 0 \\ 0 & 0 & a_1 & 0 \\ \frac{1}{2}a_4\sin(2\theta)e^{i\phi} & 0 & 0 & a_2\cos^2\theta + a_3\sin^2\theta \end{pmatrix}, \tag{5}$$

where $a_1 = 4\cosh(\eta/T)\cosh(B/T)$, $a_2 = 4e^{-\lambda J/T}\cosh^2(B/T)$, $a_3 = 4e^{\lambda J/T}\cosh^2(\eta/T)$, $a_4 = 4e^{\lambda J/T}J^2\sinh^2(\eta/T)/\eta^2$.

The degree of two-partite entanglement is measured by the concurrence $C$, which is defined as [10]

$$C = \max[2\max(\gamma_i) - \sum_{i=1}^{4}\gamma_i, 0], \tag{6}$$

where $\gamma_i(i=1,2,3,4)$ are the square roots of the eigenvalues of the matrix $R = \rho(\sigma_{1y} \otimes \sigma_{2y})\rho^*(\sigma_{1y} \otimes \sigma_{2y})$, in which $\rho$ is the density matrix and $\sigma_{ky}(k=1,2)$ are the Pauli operators of qubit $k$. The state is maximally entangled for $C=1$ and separable for $C=0$.

According to Eq. (6), the concurrence of the input and output states are given by

$$C_{in} = \sin(2\theta), \tag{7}$$

$$C_{out} = \max(\frac{a_4 C_{in} - 2a_1}{Z^2}, 0). \tag{8}$$

Eq. (8) demonstrates that when $C_{out}$ is positive, it increases linearly as $C_{in}$ increases. If $a_4 - 2a_1 \leq 0$, for which $C_{out} = 0$ even though the input state is maximally entangled, no entanglement can be teleported at all. Thus, the maximum temperature of entanglement teleportation satisfies $a_4(T) - 2a_1(T) = 0$. If the above equation does not have a positive solution, entanglement teleportation is not available at any temperature.

The calculation at finite temperature requires a numerical method, so we first focus on the characteristics at the $T=0$ limit. When $T\to 0$, we have $a_1 \sim e^{(\eta+|B|)/T}$, $a_4 \sim e^{(\lambda J+2\eta)/T}$, so in order to achieve $a_4 C_{in} - 2a_1 > 0$ we require:

$$\lambda J + \eta - |B| = \lambda J + \sqrt{J^2+b^2} - |B| > 0. \qquad (9)$$

The inequality does not hold without an inhomogeneous field for $J<0$ and $\lambda \geq 1$, which means $C_{out}$ is vanishing for all values of $B$. For other cases the critical uniform magnetic field is given by $B_c = \lambda J + \sqrt{J^2+b^2}$, beyond which entanglement teleportation is not available.

When $T>0$, Eq. (9) can be written approximately as $|B| < \lambda J + \sqrt{J^2+b^2} + T \cdot \ln(C_{in}/2)$ at low temperature, so the critical value of $B$ decreases as $C_{in}$ decreases and as $T$ increases.

The quantity $C_{out}$ as a function of $B$ and $T$ is plotted in Fig. 1. Fig. 1(a) and Fig. 1(b) correspond to $C_{in}=1$ and $C_{in}=0.3$, respectively. Due to the fact that $C_{out}$ is vanishing for $J<0$ and $\lambda=1$, only the $J>0$ case is demonstrated here. For $T=0$, when the uniform magnetic field is within the critical value, which is independent of $C_{in}$, the value of $C_{out}$ is equal to that of $C_{in}$. When the uniform magnetic field grows stronger, $C_{out}$ decreases sharply to zero. For $T>0$, the trend is similar, except that $C_{out}$ decreases to zero smoothly instead of suddenly. Moreover, the critical value of the uniform magnetic field decreases as $C_{in}$ decreases.

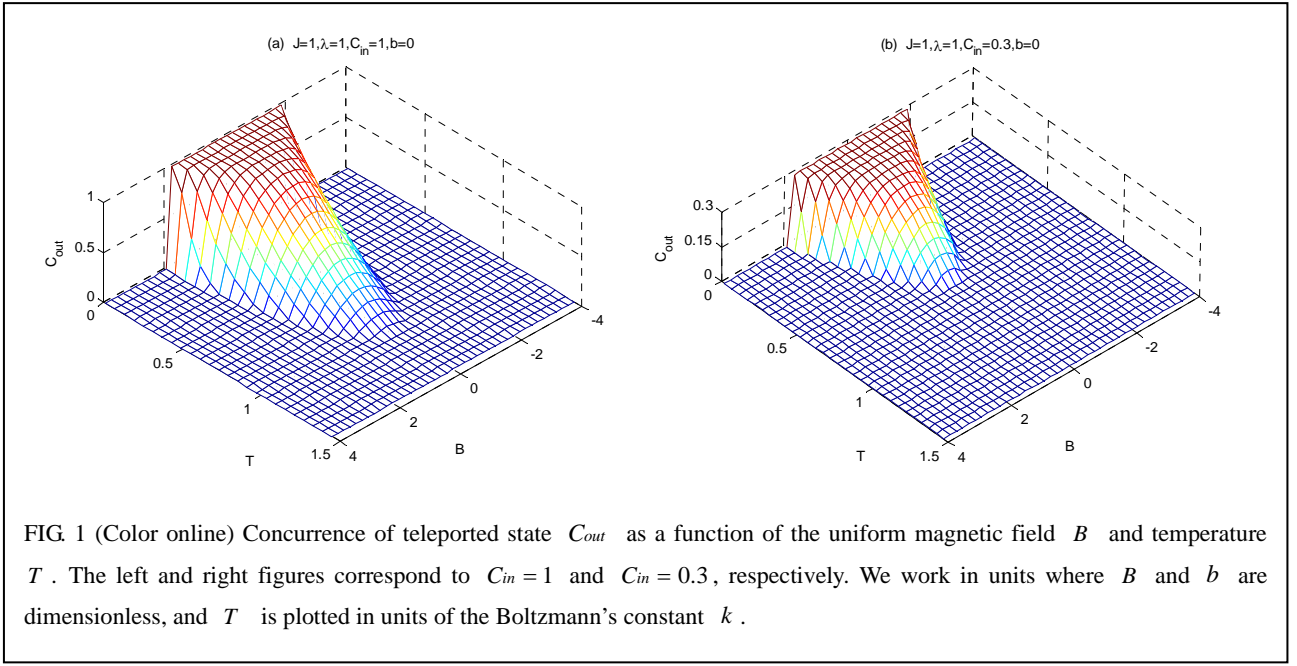

FIG. 1 (Color online) Concurrence of teleported state $C_{out}$ as a function of the uniform magnetic field $B$ and temperature $T$. The left and right figures correspond to $C_{in}=1$ and $C_{in}=0.3$, respectively. We work in units where $B$ and $b$ are dimensionless, and $T$ is plotted in units of the Boltzmann's constant $k$.

Viewed physically, at the $T=0$ limit, the entanglement of the quantum channel depends wholly on the ground state of Hamiltonian (1). Of the four eigenstates, $\varphi_3$ and $\varphi_4$ are entangled states while $\varphi_1$ and $\varphi_2$ are not. When the uniform magnetic field is weak $\varphi_3$ is the ground state. When the uniform magnetic field grows beyond the critical value, either $\varphi_1$ or $\varphi_2$ becomes the ground state, so that the channel becomes disentangled and entanglement teleportation is infeasible. In addition, the concurrence of $\varphi_3$ is independent of $B$, so when $\varphi_3$ acts as the ground state, $C_{out}$ does not change with $B$.

At finite temperature, not only the ground state, but also the eigenstates correspond to higher energy take effect in the thermal equilibrium state. When $|B|$ increases, either $E_1$ or $E_2$ decreases, so either of the two disentangled eigenstates is more distributed in the equilibrium state. As a result, $C_{out}$ decreases as $|B|$ increases at finite temperature.

Next, we turn to the effect of the inhomogeneous magnetic field. Fig. 2 shows the dependence of $C_{out}$ on $T$ and $b$ and Fig. 2(a) and Fig. 2(b) correspond to antiferromagnetic and ferromagnetic case, respectively. When $J > 0$ and the temperature is low, $C_{out}$ decreases monotonously as $|b|$ increases; at higher temperature, $C_{out}$ reaches its maximum when $|b|$ is a finite value. Thus, as the temperature rises, the $C_{out} - b$ curve transforms into a double-peaked structure, though it is not obvious in the scale of Fig. 2(a). In the $J < 0$ case, $C_{out}$ is vanishing for $b = 0$. At the $T = 0$ limit, when $|b|$ grows beyond a critical value, $C_{out}$ rises quickly to a certain value and then decreases monotonously as $|b|$ increases. The figure at higher temperature is similar, but the step is not as steep and the critical value of $|b|$ is larger.

The characteristics shown in Fig. 2 at the $T = 0$ limit can also be explained by Eq. (9). For the case in Fig. 2(a), the inequality is tenable no matter what the value of $|b|$ is, so the inhomogeneous magnetic field does not qualitatively affect the entanglement teleportation. For the case in Fig. 2(b), the critical value of $|b|$ is given by $b_c = \sqrt{(|B| - \lambda J)^2 - J^2}$, above which $\varphi_3$ becomes the ground state and both the quantum channel and the output states turn to be entangled.

The inhomogeneous magnetic field dominates the energy and the concurrence of $\varphi_3$ directly, so when $\varphi_3$ acts as the ground state, there are two effects. First, the concurrence of $\varphi_3$ decreases monotonously as $|b|$ increases. This decrease makes the concurrence of the quantum channel decrease, so $C_{out}$ generally decreases as $|b|$ increases. Second, the value of $E_3$ also decreases monotonously as $|b|$ increases, it follows that $\varphi_3$ is more distributed in the thermal equilibrium state at finite temperature, which is beneficial to entanglement teleportation. Though this effect is invisible at low temperature, it may become apparent as $T$ increases, which explains why the $C_{out} - b$ curve transforms into a double-peaked structure at higher temperature in Fig. 2(a).

Previous studies [2-4, 11] argued that a nonzero critical value of minimum thermal entanglement is required to teleport quantum entanglement. However, our study shows that this conclusion is not universal. Eq. (8) and Eq. (9) show that at zero temperature, if $|b|$ is beyond the critical value $b_c$, then $C_{out} > 0$. Thus, for an arbitrarily strong inhomogeneous magnetic field, the concurrence of output states is, though trivial, not vanishing. On the other hand, the concurrence of the quantum channel tends to zero when $|b|$ tends to infinity. Hence, though entanglement teleportation does require the quantum channel to be entangled, a nonzero critical value of minimum entanglement is not always necessary.

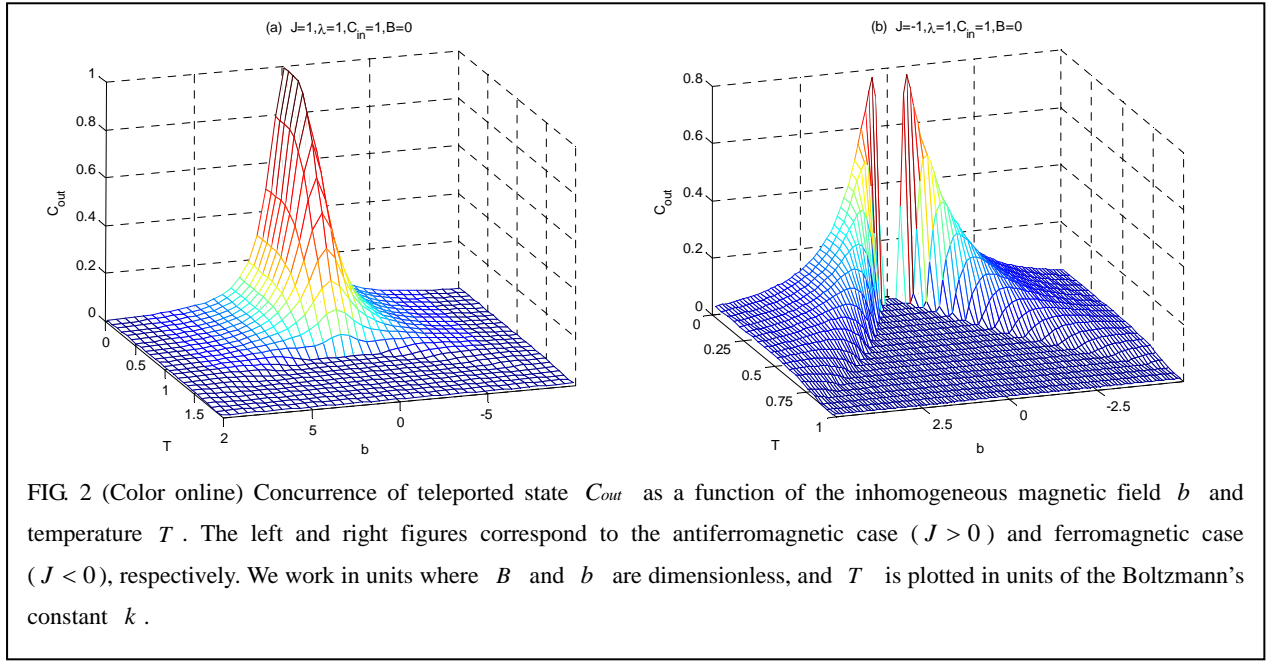

FIG. 2 (Color online) Concurrence of teleported state $C_{out}$ as a function of the inhomogeneous magnetic field $b$ and temperature $T$. The left and right figures correspond to the antiferromagnetic case ($J > 0$) and ferromagnetic case ($J < 0$), respectively. We work in units where $B$ and $b$ are dimensionless, and $T$ is plotted in units of the Boltzmann's constant $k$.

## 4. THE FIDELITY OF TELEPORTATION

In our discussion above, we mainly focused on how much entanglement is teleported. We now turn to the quality of the teleportation. In order to characterize the quality of the teleportation, it is useful to introduce the concept of fidelity [12]:

$$F(\rho_{in}, \rho_{out}) = [tr\sqrt{(\rho_{in})^{1/2} \rho_{out} (\rho_{in})^{1/2}}]^2. \tag{10}$$

The fidelity is essentially the measurement of distance between two quantum states. The two states are equal for $F = 1$ and orthogonal for $F = 0$. Thus, the larger $F$ is, the better the teleportation is achieved. In our model $F$ is expressed as

$$F(\rho_{in}, \rho_{out}) = \frac{1}{Z^2}\left[a_3 + \frac{1}{2}(a_2 - a_3 + a_4)\sin^2(2\theta)\right]. \tag{11}$$

Generally speaking, the behavior of $F$ is similar to that of $C_{out}$, except that it also depends on the input states. Again we first focus on the characteristics at zero temperature. In case of $J > 0$, for $b = 0$, when $|B|$ is within the critical value, we have $F = 1$, and otherwise we have $F = \frac{1}{2}\sin^2(2\theta)$, as shown in Fig. 3. For a fixed $|B| < B_c$, as $|b|$ increases, $F$ decreases monotonously from 1 to $1 - \frac{1}{2}\sin^2(2\theta)$, which is shown in Fig. 4(a). When $J < 0$ and $|b|$ is smaller than the critical value, $F = \frac{1}{2}\sin^2(2\theta)$; as $|b|$ increases, $F$ first jumps to a value within the range (0.5, 1), and then tends to $1-\frac{1}{2}\sin^2(2\theta)$ monotonously, as shown in Fig. 4(b).

Fig. 3 and Fig. 4 indicate that the behavior of $F$ at finite temperature is also close to that of $C_{out}$, and we do not need to repeat our description. When the temperature is very high, the output states tend to equal mixture of the standard basis, so the fidelity tends to 0.25 independent of $\lambda$, $B$ and $b$.

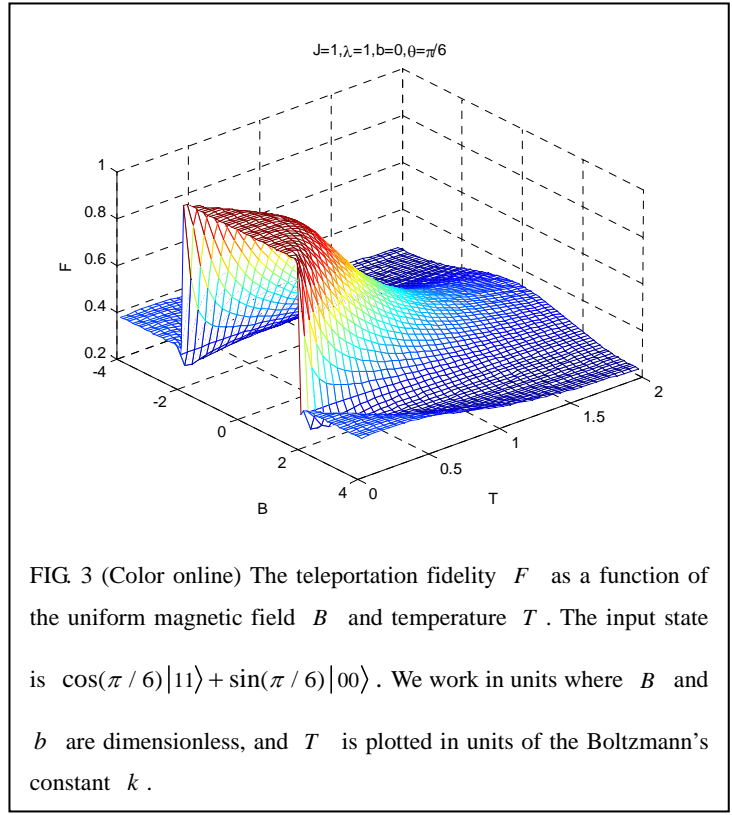

FIG. 3 (Color online) The teleportation fidelity $F$ as a function of the uniform magnetic field $B$ and temperature $T$. The input state is $\cos(\pi/6)|11\rangle + \sin(\pi/6)|00\rangle$. We work in units where $B$ and $b$ are dimensionless, and $T$ is plotted in units of the Boltzmann's constant $k$.

A remarkable feature of the fidelity is that its extremum largely ties to the value of $\theta$. To help explain this, we must evaluate Eq. (5) directly. Suppose $F=1$, then we have $a_1 = a_2 = 0$ and $a_3 = a_4 = 1$. For the case that $T=0$ and $b=0$, when $|B|$ increases beyond the critical value, both $a_3$ and $a_4$ drop from $1$ to $0$ and $a_2$ jumps from $0$ to $1$. Compared to the input state, the off-diagonal elements are vanishing and the amplitudes of $|11\rangle$ and $|00\rangle$ (i.e., also the amplitudes of $|10\rangle$ and $|01\rangle$, which are not able to emerge here) exchange, so that for most input states, the teleportation is infeasible.

As for the effect of the inhomogeneous magnetic field, if $\varphi_3$ acts as the ground state for $b=0$, then at zero temperature, the off-diagonal elements of the output state decrease to $0$ monotonously and gently as $|b|$ increases. Meanwhile the diagonal elements, which accord with the input states, are not affected. If $\varphi_3$ is not the ground state for $b=0$, then when $|b|$ increases beyond the critical value, the diagonal elements jump to the values in accord with those of the input state. Likewise, $a_4$ also jumps to nonzero values, though less than $1$. Because $a_4 = 1$ only when the magnetic field is uniform, the off-diagonal elements will be lost to some extent inevitably. Nevertheless, if $b_c$ is small, the fidelity can still exceed 2/3, which is the maximum reached by classical communication protocol. Hence, when $J<0$, the introduction of inhomogeneous magnetic field may be of benefit for the teleportation.

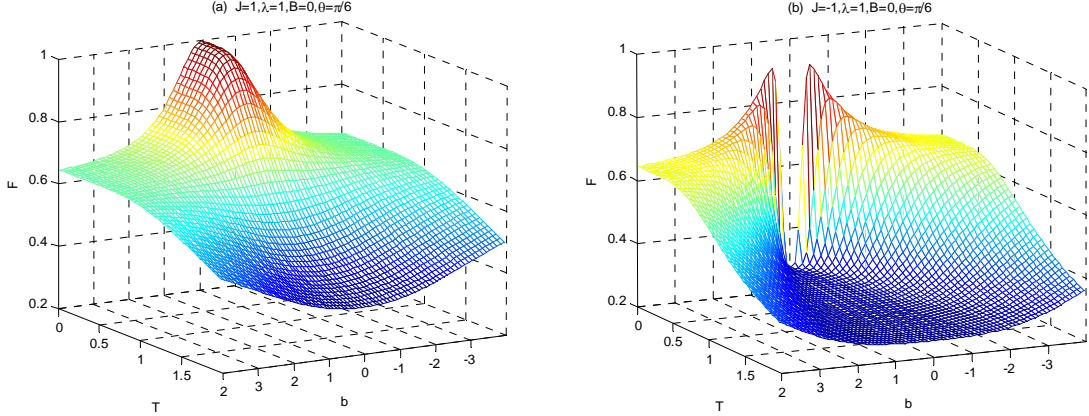

FIG. 4 (Color online) The teleportation fidelity $F$ as a function of the inhomogeneous magnetic field $b$ and temperature $T$. The left and right figures correspond to the antiferromagnetic case ($J>0$) and ferromagnetic case ($J<0$), respectively. The input state is $\cos(\pi/6)|11\rangle + \sin(\pi/6)|00\rangle$. We work in units where $B$ and $b$ are dimensionless, and $T$ is plotted in units of the Boltzmann's constant $k$.

## 5. CONCLUSION

We have investigated quantum teleportation via a two-qubit Heisenberg *XXZ* spin chain under an inhomogeneous magnetic field. We find that there is a critical value $B_c$ for the uniform magnetic field $B$. When $|B|<B_c$ it does not have much effect on quantum teleportation; otherwise the teleportation is infeasible for most input states. There are two effects of the inhomogeneous magnetic field $b$: on one hand, it decreases the teleported entanglement and fidelity generally; on the other hand, it may stimulate the entanglement of the quantum channel and enable entanglement teleportation in some ferromagnetic cases. The anisotropy parameter $\lambda$ affects the critical values of both $B$ and $b$, and it also determines whether the teleportation is feasible without an inhomogeneous magnetic field for ferromagnetic cases.

## ACKNOWLEDGEMENT

This work is supported by the National Science Foundation of China under Grant No. 10604053 and the Beihang Lantian Project.